\documentclass[conference]{IEEEtran}
\IEEEoverridecommandlockouts

\def\refname{References}
\usepackage{hyperref}
\usepackage{comment}
\usepackage{cite}
\usepackage{amsmath,amssymb,amsfonts}
\usepackage{algorithmic}
\usepackage{subcaption}
\usepackage{graphicx}
\usepackage{textcomp}
\usepackage{xcolor}
\def\BibTeX{{\rm B\kern-.05em{\sc i\kern-.025em b}\kern-.08em
    T\kern-.1667em\lower.7ex\hbox{E}\kern-.125emX}}
    
\usepackage{todonotes}
\newcommand{\linebreakand}{
  \begin{@IEEEauthorhalign}
  \hfill\mbox{}\par
  \mbox{}\hfill\end{@IEEEauthorhalign}
}
\UseRawInputEncoding
\begin{document}

\title{Performance of 802.11be Wi-Fi 7 with Multi-Link Operation on AR Applications}

\author{\IEEEauthorblockN{Molham Alsakati}
\IEEEauthorblockA{Department of Intelligent Systems\\
KTH Royal Institute of Technology, Sweden \\
}
\and
\IEEEauthorblockN{Charlie Pettersson}
\IEEEauthorblockA{Ericsson Research \\
Stockholm, Sweden \\
charlie.pettersson@ericsson.com}
\and
\IEEEauthorblockN{Sebastian Max}
\IEEEauthorblockA{Ericsson Research \\
Aachen, Germany \\
sebastian.max@ericsson.com}
\linebreakand
\and
\IEEEauthorblockN{James Gross}
\IEEEauthorblockA{Department of Intelligent Systems \\
KTH Royal Institute of Technology, Sweden \\
}
\and
\IEEEauthorblockN{Vishnu Narayanan Moothedath}
\IEEEauthorblockA{Department of Intelligent Systems \\
KTH Royal Institute of Technology, Sweden \\
}

}
\maketitle
\begin{abstract}

Since its first release in the late 1990s, Wi-Fi has been updated to keep up with evolving user needs. Recently, Wi-Fi and other radio access technologies have been pushed to their edge when serving Augmented Reality (AR) applications. AR applications require high throughput, low latency, and high reliability to ensure a high-quality user experience. The 802.11be amendment -- which will be marketed as Wi-Fi 7 -- introduces several features that aim to enhance its capabilities to support challenging applications like {AR}. One of the main features introduced in this amendment is Multi-Link Operation (MLO) which allows nodes to transmit and receive over multiple links concurrently. When using {MLO}, traffic is distributed among links using an implementation-specific traffic-to-link allocation policy. This paper aims to evaluate the performance of {MLO}, using different policies, in serving {AR} applications compared to Single-Link ({SL}). Experimental simulations using an event-based Wi-Fi simulator have been conducted. Our results show the general superiority of {MLO} when serving {AR} applications. {MLO} achieves lower latency and serves a higher number of {AR} users compared to {SL} with the same frequency resources. In addition, increasing the number of links can improve the performance of MLO. Regarding traffic-to-link allocation policies, we found that policies can be more susceptible to channel blocking, resulting in possible performance degradation. 
\end{abstract}

\begin{IEEEkeywords}
Wi-Fi 7, IEEE 802.11be, MLO, Multi-link, AR 
\end{IEEEkeywords}

\section{Introduction}

The {IEEE} 802.11 standard has been improving by amending new versions to catch up on increasing user requirements. Recently, Wi-Fi and other radio access technologies have been pushed to their limit again when it comes to Extended Reality (XR) applications which is an umbrella term that refers to: Virtual Reality (VR), AR, and Mixed Reality (MR)\cite{XR-general}.

Task Group be (TGbe), who is working on the 802.11be amendment, has been developing new features with the primary goal to enhance Wi-Fi with the capabilities to achieve Extremely High Throughput (EHT) to support high-reliability and low-latency applications -- like AR. The 11be amendment introduces {MLO}, which gives  Access Points (APs) and Stations (STAs) the capability to simultaneously transfer data belonging to the same traffic flow through multiple radio interfaces \cite{11be_standards}. Wi-Fi devices using {MLO} are expected to achieve higher peak throughput, lower latency, and higher reliability \cite{MLO-general}. When using {MLO}, traffic flows at the transmitter will be distributed to different links using a predefined traffic-to-link allocation policy, which plays an essential role in the performance of {MLO} \cite{MLO-traffic}. Therefore, evaluating the performance of {MLO} using different policies is desired. 


Since the introduction of {MLO} in Wi-Fi standards, studies have investigated its effect on  Wi-Fi performance. Multiple traffic-to-link allocation policies have been evaluated in \cite{MLO-traffic} and \cite{MLO-traffic-2}. One of the evaluated policies from these works allocates all incoming flows to the less congested link. Another one balanced the load uniformly between multiple links. The last policy implemented in these works allocates traffic according to the congestion level of the links. The authors argued that the performance of {MLO} mainly depends on the traffic-to-link allocation policy implemented. The results showed that congestion-aware policies were more flexible in adapting to the state of the network. They also concluded that allocating traffic to the least congested link, instead of proportionally distributing it among links, has less traffic exposure and simplifies the procedure complexity of the policy. 

Lopez-Raventos et al. \cite{MLO-dynamic} compared the performance of dynamic and non-dynamic traffic-to-link allocation policies that depend on the congestion of channels. In the non-dynamic policies, the load is adjusted according to the congestion level and channel occupancy of channels at the receiving nodes. The congestion level is calculated only upon a data flow arrival. On the other hand, the dynamic policy updates the load adjustment periodically and upon a data flow arrival. Two policies mentioned in \cite{MLO-traffic}, \cite{MLO-traffic-2}, and \cite{MLO-dynamic} are used in this paper. Namely, the policies balancing the load among links uniformly and according to the link congestion level. However, we calculate the congestion level using a moving average and introduce two new policies, a dynamic and a non-dynamic one.

The effect of increasing the number of aggregated links on latency is investigated in \cite{MLO-latency-evaluation}. According to the study, using three links instead of one decreased the 90\textsuperscript{th} percentile latency by $93{\%}$. However, the authors noted that using five links instead of three reduced the 90\textsuperscript{th} percentile latency by only $50{\%}$. A similar investigation of different link configuration effects on MLO is studied in this paper, but we also use different policies and impose AR traffic requirements.

This paper evaluates MLO performance, supporting AR, with different link configurations and policies. To the best of our knowledge, this paper is novel regarding evaluating MLO with AR traffic. The contributions of this paper can be summarized as follows:
\begin{itemize}
    \item We evaluate and compare Wi-Fi performance using SL and MLO, with different policies, when serving AR traffic. 
    \item We study the effects of different link configurations on the delay values while imposing the AR traffic requirements.
    \item We estimate how well Wi-Fi can serve AR applications by investigating the maximum number of AR STAs that can be supported. 
    \item We show and explain the vulnerability of some policies to channel blocking.
\end{itemize}

The remainder of this paper is organized as follows: Section~\ref{background} describes the main aspects of AR applications and MLO,
descriptions of traffic-to-link allocation policies are presented in Section~\ref{policies}, Section~\ref{simulations} describes simulation setup, evaluation methodology, and results. Discussion of those results may be found in Section~\ref{Discussion}. Lastly, Section~\ref{conclusions} gives conclusion remarks.

\section{Background}
\label{background}
\subsection{AR Applications}

 Descriptive traffic models for AR applications are needed to investigate and simulate scenarios. In this paper, AR traffic models described in the 3GPP technical report for XR applications \cite{XR-models} are used as base models. The report describes different models for AR, which vary in complexity. According to the 3GPP XR traffic models, AR traffic flows can be modeled as a combination of a \emph{ Down-Link (DL) video stream}, an \emph{ Up-Link (UL) video stream}, and an \emph{UL pose/data stream}, as illustrated in Fig.~\ref{fig:AR-traffic}.
 The traffic streams implemented in this paper, using 3GPP XR traffic models \cite{XR-models} as  bases,  are described as follows:

\begin{figure}[t]
    \centering
    \includegraphics[width=\linewidth]{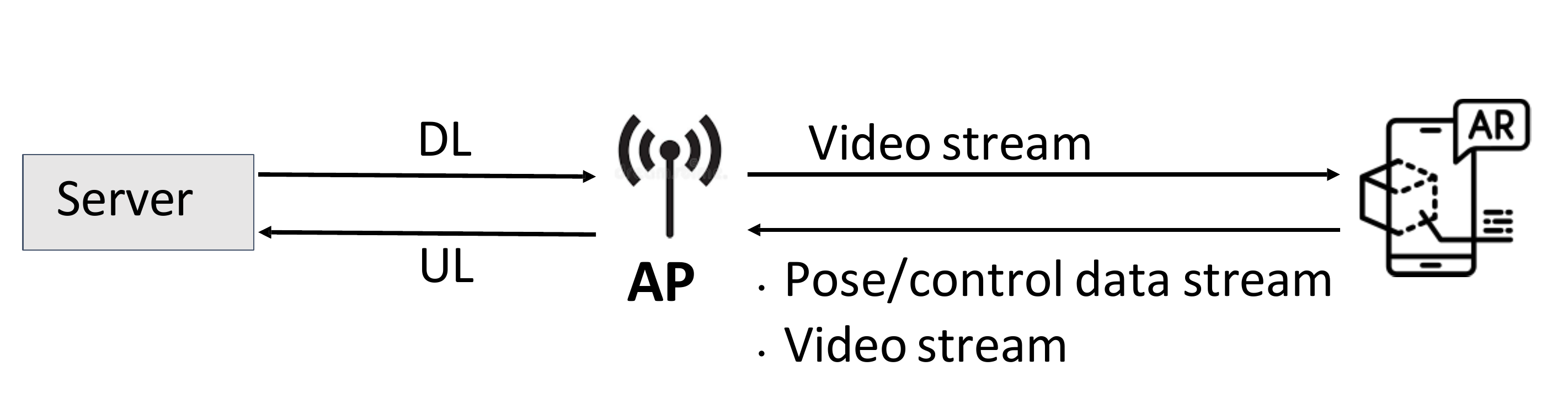}
    \caption{Illustration of AR traffic flows.} 
    \label{fig:AR-traffic}
\end{figure}

\begin{enumerate}
    \item \textbf{Video stream model:}
 This model describes the DL and UL video stream. Traffic in this model is periodic, and the packet size is modeled as a Truncated Gaussian Distribution. The network jitter is also modeled as a Truncated Gaussian Distribution in the \emph{ DL video stream}. While in the \emph{ UL video stream}, network jitter is not considered since the network distance is short. The parameter values of this model are shown in Table~\ref{tab:video_stream_model}. The data rates used in our simulations for this model were reduced to achieve higher granularity in the results. When the simulations were performed using higher data rates, as in \cite{XR-models}, many resulting Key Performance Indicator ({KPI}) values for different policies were similar, which made the evaluation comparison among policies difficult. Decreasing the data rates made the performance differences among policies clearer.
 \item \textbf{UL pose/control data stream model:}
 This model describes the traffic for \emph{pose} data needed to update the positioning and orientation of the AR user and any other \emph{control} data. Traffic from this model has a constant packet size and a fixed periodicity. The parameter values for this model can be seen in Table~\ref{tab:pose_stream_model}.
\end{enumerate}

 \begin{table}[t]
    \centering
    \small
    \begin{tabular}{|p{0.2\textwidth}|p{0.2\textwidth}|} 
    \hline
      \textbf{Parameter} & \textbf{Value}   \\
      
      \hline
      Frame rate           & $60$\,f/s            \\
      Data rate           &DL:$10$\,Mb/s UL:$3.3$\,Mb/s \\
      Periodicity       & $16.667$\,ms\\
      PDB         &DL:$10$\,ms UL:$30$\,ms\\
      Packet success rate    & $99$\,\% \\
      \hline
      \multicolumn{2}{|c|}{\textbf{Packet size model}} \\
      \hline
      Mean      &DL:$21$\,KB UL:$7$\,KB\\
      STD       & 10.5\% of Mean \\
      Max.       & $150$ \% of Mean                    \\
      Min.       & $50$\% of Mean\\
      \hline
      \multicolumn{2}{|c|}{\textbf{Jitter model (DL only)}} \\
      \hline
      Mean   & $0$\,ms\\
      STD & $2$\,ms\\
      Max.       & $4$\,ms\\
      Min.       & $-4$\,ms\\
      \hline
    \end{tabular}
    \caption{DL/UL video stream parameters. }
    \label{tab:video_stream_model}
\end{table}

\begin{table}[t]
\centering   \small 
    \begin{tabular}{|p{0.2\textwidth}|p{0.2\textwidth}|} 
    \hline
      \textbf{Parameter} & \textbf{Value}   \\
      
      \hline
      Packet size           &$100$\,B   \\
      Periodicity       & $4$\,ms \\
      PDB        & $10$\,ms\\
      Packet success rate    & $99$\,\% \\
      \hline
    \end{tabular}
     \caption{UL pose/control stream parameters. }
     \label{tab:pose_stream_model}
\end{table}

These models have two primary required limits. First, the Packet Delay Boundary (PDB), which is the maximum delay allowed for application packets to be received successfully. The packet delay is an application layer delay value measured from when the packet arrives at the AP or STA to when its destination successfully receives it. The second limit is the packet success rate, which is the percentage of packets arriving within the PDB. The required packet success rate is 99\% for all three data streams.

\subsection{Multi-Link Operation}
The idea of MLO, as introduced in the 802.11be amendment, is to use multiple radio interfaces to transmit and receive data concurrently. These radio interfaces can operate simultaneously on the 2.4, 5, and 6 GHz bands \cite{MLO-general} \cite{MLO-general-2}. 

Since MLO uses multiple separate links to transmit and receive, adjustments have been suggested to control the channel access mechanisms for these links. Many transmission modes has been suggested by TGbe. This paper considers the asynchronous transmission mode since it achieves the best MLO performance \cite{MLO_legacy_and_ax}. The asynchronous transmission mode enables nodes to transmit and receive packets over multiple interfaces simultaneously.

The concept of Multi-Link Device (MLD) is also introduced in the 11be amendment. An MLD is a device with multiple Physical (PHY) interfaces sharing the same interface to the Logical Link Control (LLC) layer. In other words, MLD is a device that implements MLO. 

To implement MLO,  11be divided the Medium Access Control (MAC) layer into two sublayers \cite{MAC-layers}: A Lower MAC that is unique for each interface and supports link-specific operations like channel access, link adaptation, and sounding. The second sublayer is called the Upper MAC and is shared among interfaces. It supports operations like Aggregated MAC Service Data Unit (AMSDU) aggregation/de-aggregation,  packet numbering, and fragmentation/defragmentation. In the Upper MAC, the distribution of MAC Protocol Data Units (MPDUs) among links is conducted according to the applied traffic-to-link allocation policy, which is implementation-specific. A description of such a policy along with implemented policies definitions are presented in the next section. 

\section{Traffic-to-link allocation policies}
\label{policies}

The traffic-to-link mapping takes place in the Upper MAC using a \emph{traffic-to-link allocation policy}. When data to be transmitted arrives at a MAC buffer, it is containerized in MPDUs, and MLD transmitters are notified to start their independent channel access processes. Depending on the policy used, the Upper MAC layer allocates a specific number of MPDUs from the buffer to be transmitted by each link. 

Policies presented in this paper are divided into \emph{Informed} and \emph{Uninformed} policies. The \emph{Uniformed} policies are straightforward and independent of the real-time system state. They consist of the \emph{Greedy} and \emph{Uniform-Load} policies. In contrast, the \emph{Informed} policies are dynamic and consider system information when distributing traffic among links. The \emph{Informed} policies in this paper are the \emph{Congestion-Aware} and \emph{Condition-Aware} policies. The implemented policies are defined as follows:

\begin{enumerate}
    \item \textbf{Greedy Policy}:
    This policy allocates the maximum allowed number of MPDUs from the MAC buffer to the first link that wins access to the medium.
    
    \item \textbf{Uniform-Load Policy}:
    This policy distributes the MPDUs present in the MAC buffer uniformly among radio links. Assume that MLO uses $i$ links and $n$ MPDUs are pending in the MAC buffer. An AMPDU consisting of $\lceil n/i \rceil$ MPDUs is aggregated and allocated for each link.
    
    \item \textbf{Congestion-Aware Policy}:
    This policy aims to balance the traffic load by utilizing estimated congestion levels for each link. The congestion level on a link can be described using a \emph{channel busy-time} metric, which is the fraction of a particular period that a channel is busy \cite{congestion_definition}. This particular period is called the \emph{Update Period}.
    
    In our simulations, the estimated congestion level depends on the transmission duration for all received packets on a link during the update period. During an update period, each link on each node accumulates the transmission duration of all received packets, including packets that do not belong to the node. 
    
    The estimated congestion level, measured as \emph{Link busy-time},  is then updated using a moving average.
    A window containing the ten previous congestion level values is used for the moving average. The choice of update period and moving average window values was decided experimentally. First, the theoretical congestion level is calculated on each link in a test setup. Then, the estimated congestion levels produced by this policy are checked. The chosen values of 0.5\,s and 10 samples for the update period and the moving average window, respectively, give a close enough estimation for the congestion level. 
    
    The policy distributes a ratio of packets from the buffer to a link  according to the following equation: 
    \[
	\text{Packet ratio} = \frac{\text{Link free time}}{\text{Total free time}}
\] 
    \emph{Total free time} is the sum of \emph{Link free time} values from all links. The free time on each link is calculated as shown below:
        \[
   \text{Link free time} =  \text{Update period}- \text{Link busy time}
\] 

    \item \textbf{Condition-Aware Policy}:
    Since there can be links with high congestion but can transmit data with higher data rates, a policy that considers the transmission data rate could be beneficial. This policy is similar to the Congestion-aware policy but also considers the effect of the subsequent transmission data rate for each link. 
    
    To calculate the ratio of packets for each link, the \emph{Information bits per period} parameter is introduced. This parameter corresponds to the estimated number of information bits a link can transmit in a congestion update period (0.5 s as in the Congestion-aware policy). The \emph{information bits per period} for each link is calculated as follows:
\[
	\text{Information bits per period} = \text{Link free time} \cdot \text{Data rate}
\]
The \emph{Data rate} is calculated, for each link, using the link bandwidth and the decided Modulation and Coding Scheme (MCS) value of the subsequent transmission. The MCS value is decided according to the link adaptation algorithm used, namely the Minstrel adaptation. The ratio of packets allocated for each link is then found as below:
\[
	\text{Packets ratio} = \frac{\text{Information bits per period}}{\text{Total information bits per period}}
\] 
Where the \emph{Total information bits per period} is the sum of \emph{Information bits per period} from all links. 
    \end{enumerate}
All the aforementioned policies are implemented in our Wi-Fi simulator and used in each simulation setup throughout the paper. Simulation details and results are presented in the next section.
\section{Simulations}
\label{simulations}
Details regarding the simulation scenario, Wi-Fi simulator, evaluation methodology, and obtained results are discussed in this section.

\subsection{Methodology}
\begin{figure}[ht]
    \centering
    \includegraphics[width=\linewidth]{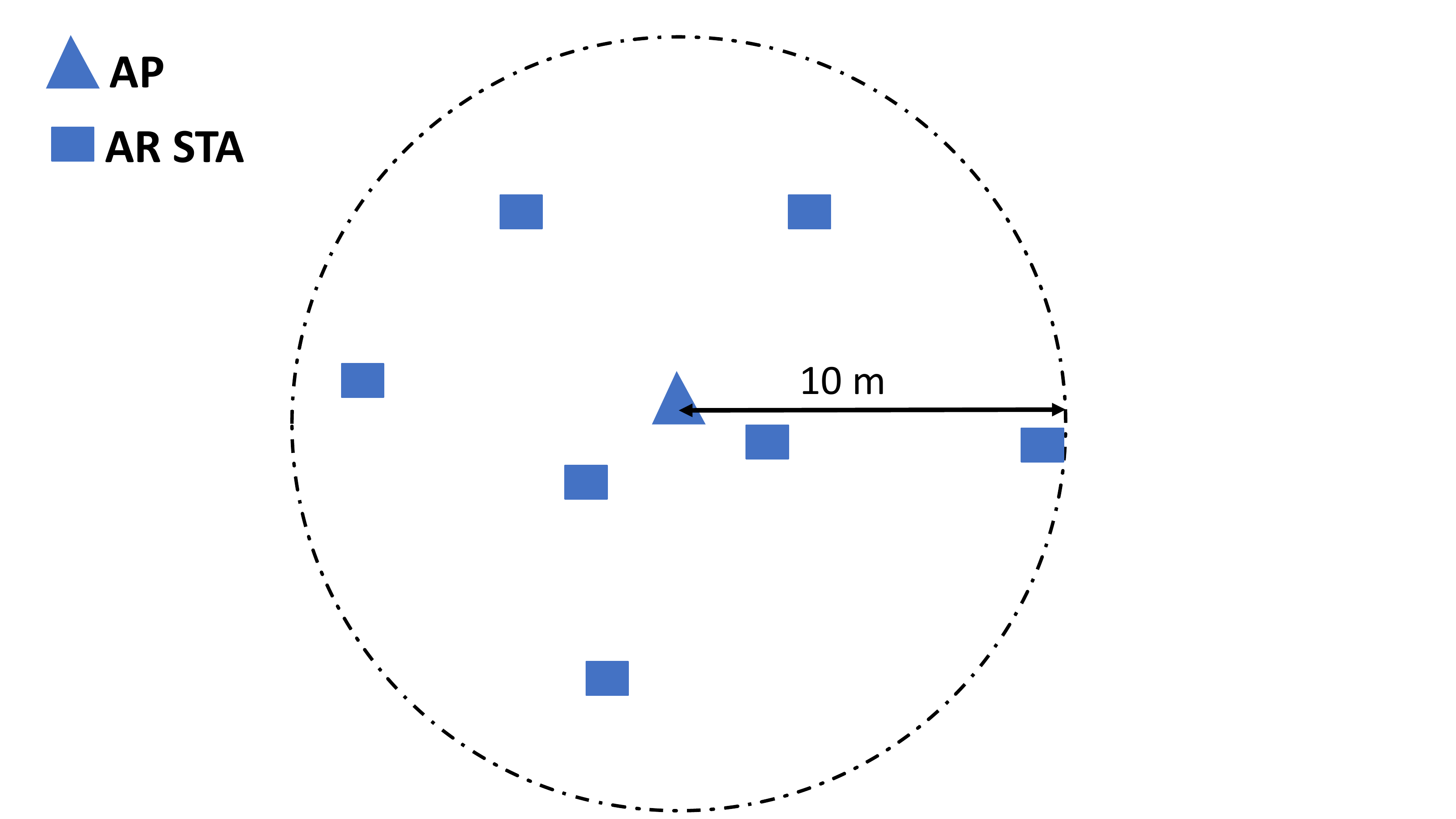}
    \caption[Single-cell scenario deployment]{The AP and STAs deployment in the single-cell scenario.} 
    \label{fig:single-cell_deployment}
\end{figure}

In this paper, we consider a single-cell scenario that consists of one AP in the center and several STAs that spawn at random positions with a maximum distance of 10 m to the AP, as shown in Fig.~\ref{fig:single-cell_deployment}. A proprietary, well-established, event-based Wi-Fi system simulator was used to implement the designed scenario and run simulations. The simulator has been used in research for many years and is in continuous development. In our simulations, all STAs are activated within the first second of the simulation and remain active throughout the simulation time of 50 seconds. For each configuration simulations with different seeds are performed to ensure statistical relevance. System parameters used in this scenario are presented in Table~\ref{tab:system_parameters}.

 \begin{table}[t]
    \centering
    \small
    \begin{tabular}{|p{0.28\textwidth}|p{0.16\textwidth}|} 
    \hline
      \textbf{Parameter} & \textbf{Value}   \\
       \hline
    \small Max. AMPDU length & 5.484\,ms\\
    
    \small Max. number of MPDUs in an AMPDU &64 MPDUs\\
    
    MCS selection algorithm &  Minstrel\\

    Transmit power & 20\,dBm\\
   
    MPDU payload size & 1500\,B\\
     
    Wireless channel model & TGn model D \cite{tgnd}\\
    \hline
    \multicolumn{2}{|c|}{\textbf{Carrier frequencies}} \\\hline
    SL  & 5.5\,GHz\\
   
    MLO  & 5.2, 5.5, 6.1, and 6.5\,GHz\\
      \hline
      \multicolumn{2}{|c|}{\textbf{Link bandwidth(s)}} \\\hline
    SL  &  80 and 160\,MHz\\
     
    MLO  &2$\times$40, 4$\times$20, and 2$\times$80\,MHz\\
      \hline
    \end{tabular}
    \caption{System Parameters. }
    \label{tab:system_parameters}
\end{table}

The main KPI we are interested in evaluating is the maximum number of STAs that Wi-Fi can support, considering the AR traffic requirements. Therefore, we log the delay values for each STA and data stream in each simulation. 

After gathering all data belonging to a data stream and an STA, an evaluation of whether the system fulfilled the application requirements is conducted. Let the number of STAs in the simulation setup be $m$. For each data stream, the 99\textsuperscript{th} percentile delay values for each STA are calculated, given that the Packet Success Rate requirement for all data streams is 99\%, as mentioned above. Then, the STA with the worst 99\textsuperscript{th} percentile delay value in each data stream is identified and compared with the corresponding delay limitations. If the worst 99\textsuperscript{th} percentile delay value in each data stream is less than or equal to the corresponding delay limit, one can deduce that the system can successfully support at least $m$ STAs in the simulated setup.

\begin{figure*}[h]
    \centering
    \includegraphics[width=\linewidth]{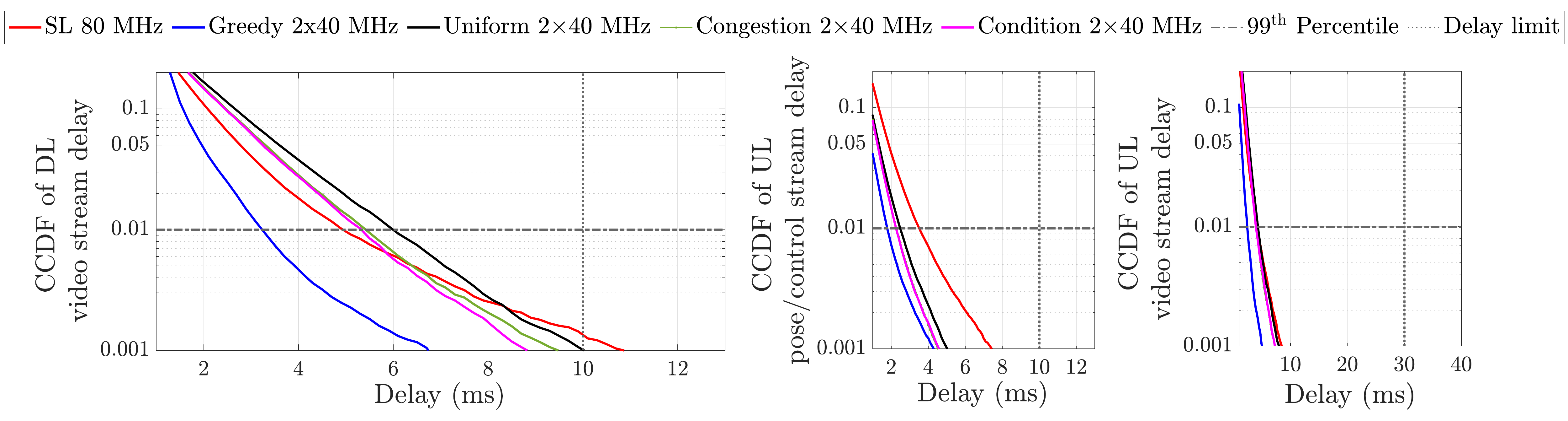}
    \caption{Resulted CCDF plots of delay values for each AR data stream when simulating with 6 STAs, using SL and different MLO policies. }
    \label{ccdf}
\end{figure*}

An illustration of such evaluation is shown in Fig~\ref{ccdf}, which contains the  Complementary Cumulative Distribution Function (CCDF) delay plots for the three data streams when simulating with 6 STAs. Note that, the CCDF should be less than 0.01 -- which corresponds to the 99\textsuperscript{th} percentile -- for all delay values higher than the delay limit. Thus, one can infer from this illustration that the Wi-Fi system can support 6 STAs using SL or any MLO policies. Furthermore, note from the figure that the DL stream generates worst-case delays that are closer to the limit than those of the two UL streams. As a result, in our simulations, it is sufficient to consider only the data corresponding to the DL stream while checking the KPI of the number of STAs that can be supported.

\subsection{Results}

\begin{figure*}[htbp]
    \centering
    \includegraphics[width=\linewidth]{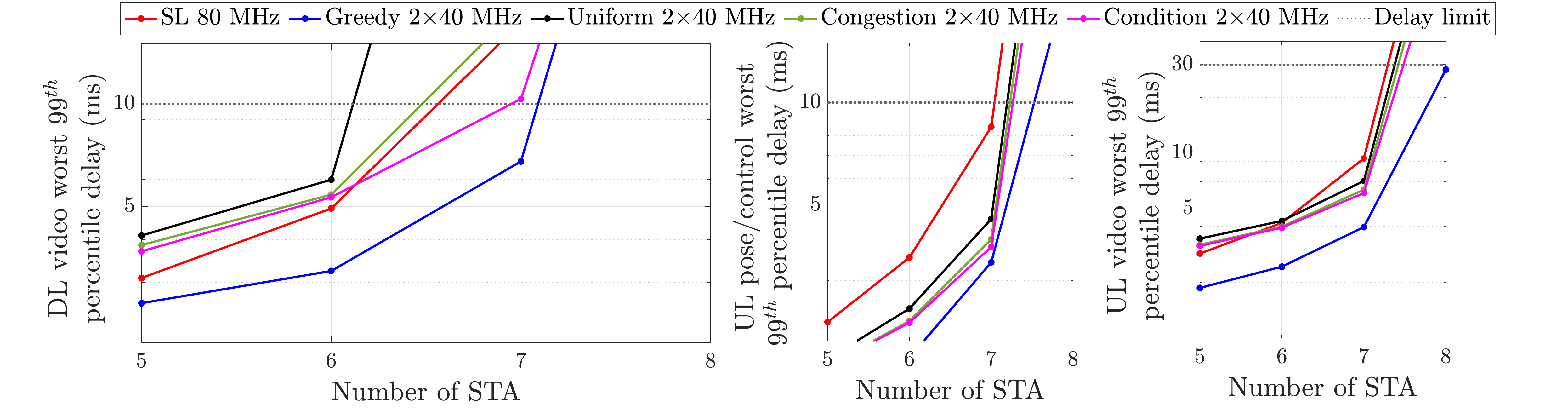}
    \caption{Worst 99\textsuperscript{th} delay values for each AR data stream with different policies plotted against the number of STAs. Link bandwidth for SL is  80\,MHz and 2$\times$40\,MHz for MLO.}
    \label{2x40}
\end{figure*}

In Fig~\ref{2x40}, we show results from simulating with different numbers of STAs and the corresponding worst 99\textsuperscript{th} percentile delay values for each data stream. The total bandwidth is chosen to be 80 MHz, that is, 80 MHz for SL, and 2$\times$40 for MLO. In the plot corresponding to the DL video delay, we can see that all policies support a maximum of 6 STAs except the Greedy policy which can support 7 STAs. As discussed earlier, note that the delay values are higher for the DL video stream than those in UL streams with the same number of STAs, thus making it the dominant figure. As a result, even though the UL streams can support 7 STAs (the UL video stream with Greedy policy can support even 8 STAs), we do not consider the system successful as the DL stream violates the delay limit.

Simulating with other link configurations, namely using 4$\times$20 and 2$\times$80 MHz, resulted in similar behavior where the DL stream violates its corresponding AR traffic requirements and is the limiting factor regarding the maximum number of supported STAs. Therefore, we choose not to show these figures to avoid redundancy. 

\begin{figure}[ht]
    \centering
    \includegraphics[width=\linewidth]{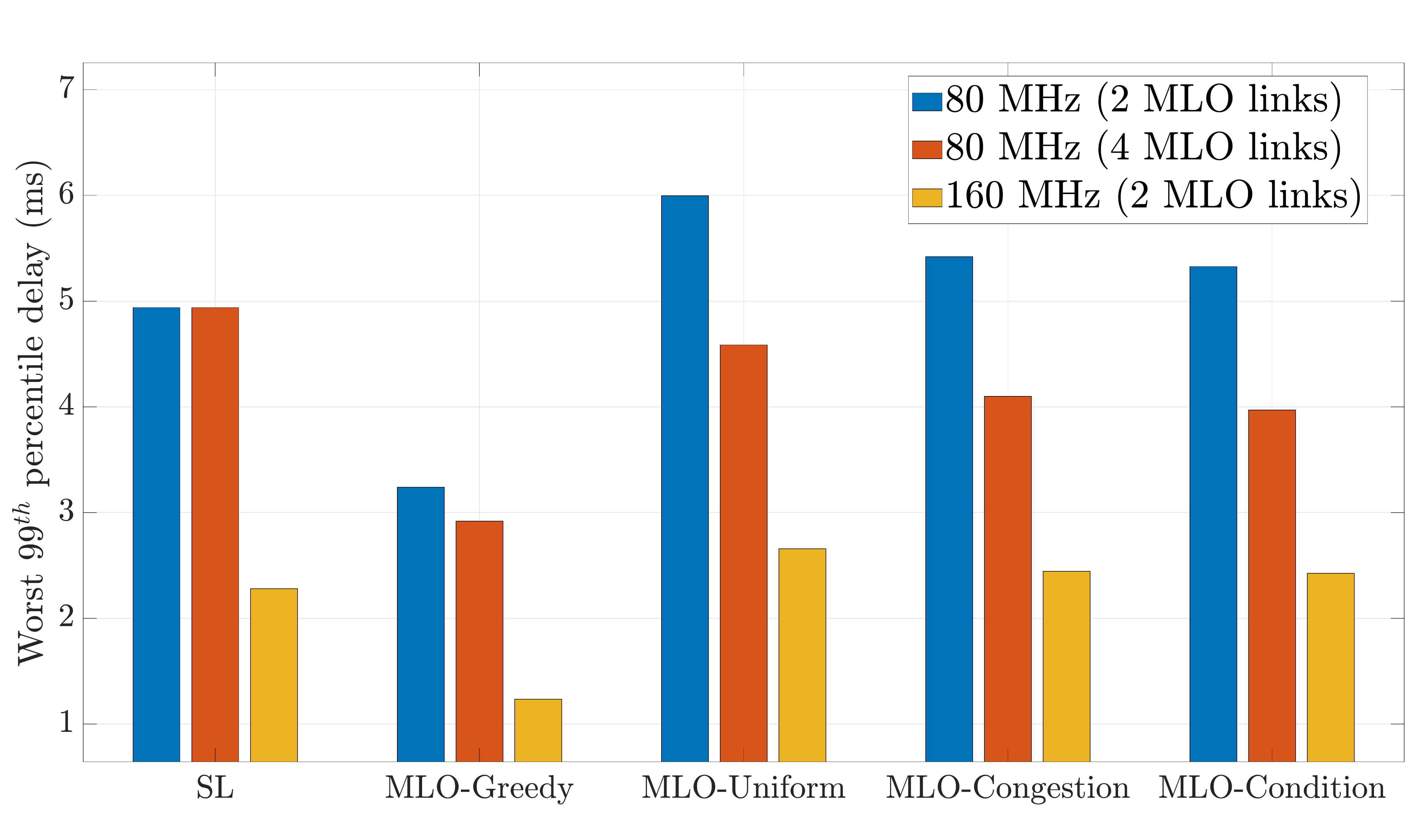}
    \caption{The worst 99\textsuperscript{th} DL delay values when simulating with 6 STAs, using SL and different
MLO policies.} 
    \label{hist:fixd_sta_vs_delay}
\end{figure}

Fig~\ref{hist:fixd_sta_vs_delay} shows an overview of 
the worst 99\textsuperscript{th} percentile delay values, belonging to the DL video stream,  when supporting 6 STAs with various link configurations and policies. This figure shows that with a link configuration of 2$\times$40\,MHz, the Uniform, Congestion, and Condition policies have higher delay values than SL, with the Greedy policy having the lowest delay value. Increasing the number of links using the same bandwidth, namely using 4 links with 20 MHz each, yields lower delay values for MLO in general. A similar effect can result when increasing the bandwidth to 160 MHz. We can conclude from this figure that the Greedy policy achieves the lowest delay values among other policies and with different link configurations.

\begin{figure}[ht]
    \centering
    \includegraphics[width=\linewidth]{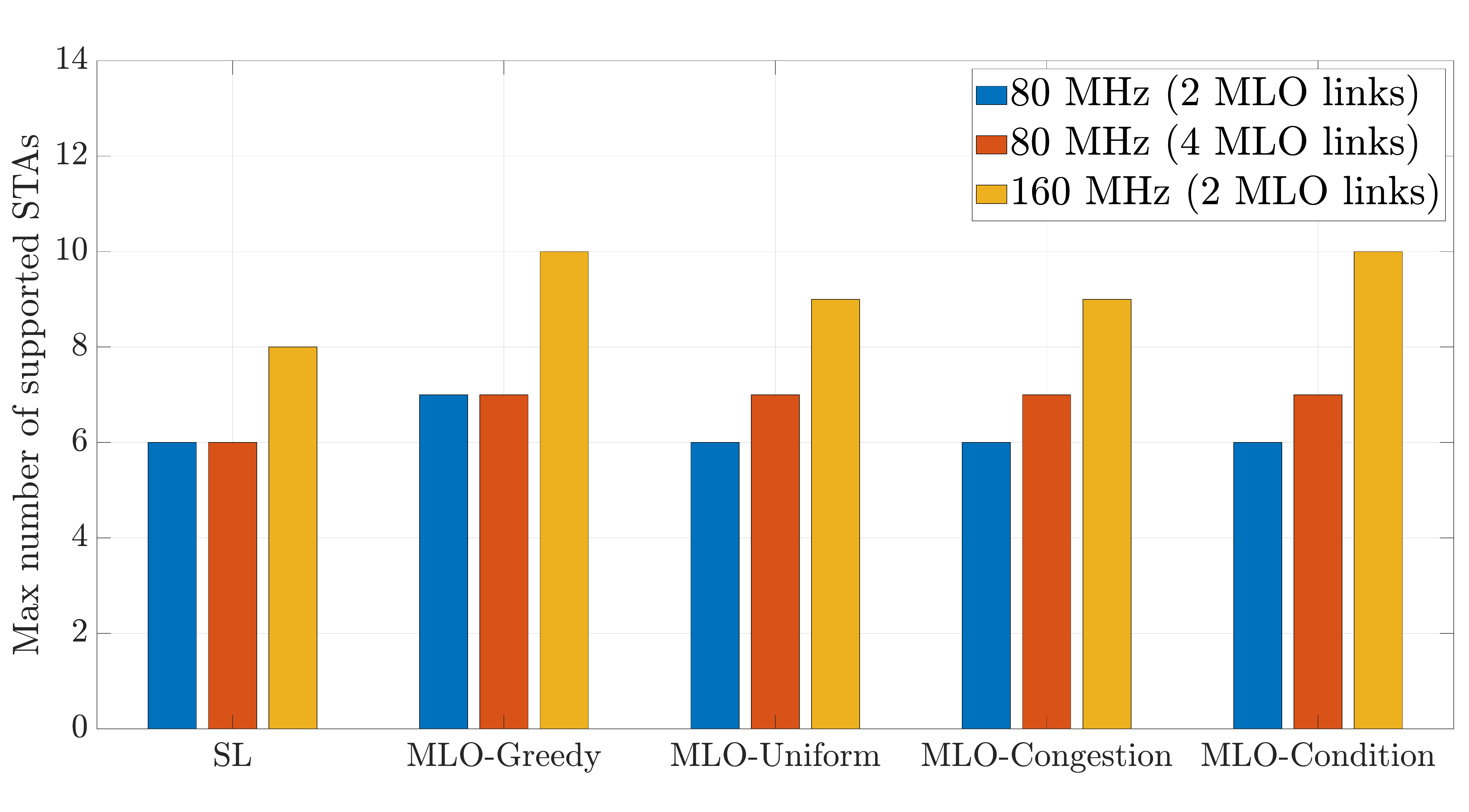}
    \caption{The maximum number of AR STAs that can be supported by each policy and link configuration.} 
    \label{hist:max_sta}
\end{figure}

Fig~\ref{hist:max_sta} shows the main evaluation KPI values for SL and MLO. From this figure, we notice that when simulating with a 2$\times$40 MHz link configuration, Greedy policy has the best performance with 7 STAs that can be supported. In comparison, using SL or MLO with the other policies can support a maximum of 6 STAs, with the same frequency resources. Increasing the number of links to four gives MLO, using any policy, a gain of around 17\% compared to SL. When simulating with a total of 160 MHz bandwidth, a maximum of 10 STAs can be supported using Greedy and Condition policies, 9 STAs with the Uniform and Congestion policies, while SL can support only 8 STAs.

\section{Discussion}
\label{Discussion}
Considering the results shown in Fig~\ref{hist:max_sta}, 
we surprisingly notice that the informed policies are performing either worse or as good as the uninformed Greedy policy. We can also see, from Fig~\ref{ccdf} and Fig~\ref{2x40}, that the worst 99\% delay values in all streams are the lowest when using the Greedy policy. This behavior is counter-intuitive since the informed policies are supposed to perform better considering their ability to adjust dynamically to the status of channels, while the Greedy policy acts statically.

Frame and packet distribution analyses were conducted for each policy to investigate the degradation in the informed policies' performance. We found that they, plus the Uniform policy, have a similar behavior regarding channel access. The implementation strategy used for these policies assumes that all links will get access to the channel in a reasonable time and transmit their allocated AMPDUs. In other words, these policies require simultaneous access to all links to perform as we expect. Therefore, we call these policies  Simultaneous-Access Policies (SAPs). 

The problem with such an implementation strategy is that there is no guarantee of getting access due to the contention-based channel access in Wi-Fi. We have observed that in some cases, it can happen that one of the links does not get access to the medium in a reasonable time. Then, the packets that were supposed to be transmitted on that link will remain in the buffer. When block acknowledgments are received at other links, the policy restarts, and the remaining packets will be distributed among links again. The same behavior can be repeated with a non-zero probability until one or two packets remain in the buffer, which can be transmitted in the same AMPDU. This behavior can be present and result in acceptable delay values. However, when the environment gets more congested, e.g. by adding STAs, the probability of collisions and channel blocking becomes higher.

Consequently, packets have delay values over the required limits due to higher queuing delays, leading to performance degradation. In the case of using the Greedy policy, only one channel access is needed to transmit the application frame using one AMPDU. Therefore, this policy does not suffer from unreliable channel access to the same extent. 

We conclude from the analysis of SAPs that these policies can make transmissions more vulnerable to channel blocking. However, MLO using a wisely-chosen policy can serve more AR STAs than SL, as shown in Fig~\ref{hist:max_sta}. MLO can achieve lower delay values and is less affected by incrementing the number of supported STAs due to parallel access and transmission. 

\section{Conclusions}
\label{conclusions}

From the results obtained, we conclude that MLO can achieve lower latency than SL when serving AR. Therefore, using the same frequency resources, MLO serves more AR STAs. However, we can conclude that informed policies do not always perform better than uninformed ones and that Greedy policy performs as good, or even better than,  informed policies. We also found that the implementation choice of the SAPs caused an issue due to unreliable channel access and degraded their performance when simulating with AR. Increasing the number of links by two, using the same frequency resources, achieved lower latency values and can yield a higher number of supported STAs.

For future work, it might be interesting to study different channel access techniques to alleviate the vulnerability of SAPs to channel blocking. 


\bibliographystyle{IEEEtran}

\bibliography{references}

\begin{thebibliography}{10}
\providecommand{\url}[1]{#1}
\csname url@samestyle\endcsname
\providecommand{\newblock}{\relax}
\providecommand{\bibinfo}[2]{#2}
\providecommand{\BIBentrySTDinterwordspacing}{\spaceskip=0pt\relax}
\providecommand{\BIBentryALTinterwordstretchfactor}{4}
\providecommand{\BIBentryALTinterwordspacing}{\spaceskip=\fontdimen2\font plus
\BIBentryALTinterwordstretchfactor\fontdimen3\font minus
  \fontdimen4\font\relax}
\providecommand{\BIBforeignlanguage}[2]{{%
\expandafter\ifx\csname l@#1\endcsname\relax
\typeout{** WARNING: IEEEtran.bst: No hyphenation pattern has been}%
\typeout{** loaded for the language `#1'. Using the pattern for}%
\typeout{** the default language instead.}%
\else
\language=\csname l@#1\endcsname
\fi
#2}}
\providecommand{\BIBdecl}{\relax}
\BIBdecl

\bibitem{XR-general}
\BIBentryALTinterwordspacing
S.~Doolani, C.~Wessels, V.~Kanal, C.~Sevastopoulos, A.~Jaiswal, H.~Nambiappan,
  and F.~Makedon, ``\BIBforeignlanguage{en}{A {Review} of {Extended} {Reality}
  ({XR}) {Technologies} for {Manufacturing} {Training}},''
  \emph{\BIBforeignlanguage{en}{Technologies}}, vol.~8, no.~4, p.~77, Dec.
  2020, number: 4 Publisher: Multidisciplinary Digital Publishing Institute.
  [Online]. Available: \url{https://www.mdpi.com/2227-7080/8/4/77}
\BIBentrySTDinterwordspacing

\bibitem{11be_standards}
E.~Au, ``{IEEE} 802.11be: {Extremely} {High} {Throughput} [{Standards}],''
  \emph{IEEE Vehicular Technology Magazine}, vol.~14, no.~3, pp. 138--140, Sep.
  2019, conference Name: IEEE Vehicular Technology Magazine.

\bibitem{MLO-general}
\BIBentryALTinterwordspacing
E.~Khorov, I.~Levitsky, and I.~F. Akyildiz, ``Current {Status} and {Directions}
  of {IEEE} 802.11be, the {Future} {Wi}-{Fi} 7,'' \emph{IEEE Access}, vol.~8,
  pp. 88\,664--88\,688, 2020. [Online]. Available:
  \url{https://ieeexplore.ieee.org/document/9090146/}
\BIBentrySTDinterwordspacing

\bibitem{MLO-traffic}
\BIBentryALTinterwordspacing
A.~Lopez-Raventos and B.~Bellalta, ``Multi-link {Operation} in {IEEE} 802.11be
  {WLANs},'' \emph{arXiv:2201.07499 [cs]}, Jan. 2022, arXiv: 2201.07499.
  [Online]. Available: \url{http://arxiv.org/abs/2201.07499}
\BIBentrySTDinterwordspacing

\bibitem{MLO-traffic-2}
\BIBentryALTinterwordspacing
------, ``{IEEE} 802.11be {Multi}-{Link} {Operation}: {When} the {Best} {Could}
  {Be} to {Use} {Only} a {Single} {Interface},'' in \emph{2021 19th
  {Mediterranean} {Communication} and {Computer} {Networking} {Conference}
  ({MedComNet})}.\hskip 1em plus 0.5em minus 0.4em\relax Ibiza, Spain: IEEE,
  Jun. 2021, pp. 1--7. [Online]. Available:
  \url{https://ieeexplore.ieee.org/document/9501237/}
\BIBentrySTDinterwordspacing

\bibitem{MLO-dynamic}
\BIBentryALTinterwordspacing
------, ``Dynamic {Traffic} {Allocation} in {IEEE} 802.11be {Multi}-link
  {WLANs},'' \emph{arXiv:2202.12614 [cs]}, Feb. 2022, arXiv: 2202.12614.
  [Online]. Available: \url{http://arxiv.org/abs/2202.12614}
\BIBentrySTDinterwordspacing

\bibitem{MLO-latency-evaluation}
G.~Naik, D.~Ogbe, and J.-M.~J. Park, ``Can {Wi}-{Fi} 7 {Support} {Real}-{Time}
  {Applications}? {On} the {Impact} of {Multi} {Link} {Aggregation} on
  {Latency},'' in \emph{{ICC} 2021 - {IEEE} {International} {Conference} on
  {Communications}}, Jun. 2021, pp. 1--6, iSSN: 1938-1883.

\bibitem{XR-models}
\BIBentryALTinterwordspacing
``Study on {XR} ({Extended} {Reality}) evaluations for {NR},'' 3GPP, Technical
  report 38.838. [Online]. Available:
  \url{https://www.3gpp.org/ftp/Specs/archive/38_series/38.838/38838-h00.zip}
\BIBentrySTDinterwordspacing

\bibitem{MLO-general-2}
\BIBentryALTinterwordspacing
C.~Deng, X.~Fang, X.~Han, X.~Wang, L.~Yan, R.~He, Y.~Long, and Y.~Guo, ``{IEEE}
  802.11be {Wi}-{Fi} 7: {New} {Challenges} and {Opportunities},'' \emph{IEEE
  Communications Surveys \& Tutorials}, vol.~22, no.~4, pp. 2136--2166, 2020.
  [Online]. Available: \url{https://ieeexplore.ieee.org/document/9152055/}
\BIBentrySTDinterwordspacing

\bibitem{MLO_legacy_and_ax}
G.~Lacalle, I.~Val, O.~Seijo, M.~Mendicute, D.~Cavalcanti, and
  J.~Perez-Ramirez, ``Analysis of {Latency} and {Reliability} {Improvement}
  with {Multi}-{Link} {Operation} over 802.11,'' in \emph{2021 {IEEE} 19th
  {International} {Conference} on {Industrial} {Informatics} ({INDIN})}, Jul.
  2021, pp. 1--7.

\bibitem{MAC-layers}
\BIBentryALTinterwordspacing
F.~Yonggang, B.~Sun, Z.~Han, and N.~Li, ``Multi-{Link} {Architecture} and
  {Requirement} {Discussion},'' Jul. 2019. [Online]. Available:
  \url{https://mentor. ieee.org/802.11/documents?is_dcn=1095&is_group=00be.}
\BIBentrySTDinterwordspacing

\bibitem{congestion_definition}
\BIBentryALTinterwordspacing
A.~P. Jardosh, K.~N. Ramachandran, K.~C. Almeroth, and E.~M. Belding-Royer,
  ``\BIBforeignlanguage{en}{Understanding congestion in {IEEE} 802.11b wireless
  networks},'' in \emph{\BIBforeignlanguage{en}{Proceedings of the 5th {ACM}
  {SIGCOMM} conference on {Internet} measurement - {IMC} '05}}.\hskip 1em plus
  0.5em minus 0.4em\relax Berkeley, California: ACM Press, 2005, p.~1.
  [Online]. Available:
  \url{http://portal.acm.org/citation.cfm?doid=1330107.1330140}
\BIBentrySTDinterwordspacing

\bibitem{tgnd}
\BIBentryALTinterwordspacing
V.~Erceg, L.~Schumacher, P.~Kyritsi, and A.~Molisch, ``{TGn} {Channel}
  {Models},'' IEEE, Tech. Rep. IEEE 802.11-03/940r4, May 2004. [Online].
  Available:
  \url{https://mentor.ieee.org/802.11/dcn/03/11-03-0940-04-000n-tgn-channel-models.doc}
\BIBentrySTDinterwordspacing

\end{thebibliography}

\end{document}